\documentclass[reprint,aps,prl]{revtex4-1}

\usepackage{graphicx}
\usepackage{bm}
\usepackage{amsmath}

\begin{document}

\title{Diffusion of Janus Particles in Bacterial Suspensions:\\
Second-Order Phase Transition and Counterintuitive Directional Propulsion}

\author{Zihan Huang}
\author{Pengyu Chen}
\author{Guolong Zhu}
\author{Yufei Cao}
\author{Li-Tang Yan}
 \email{ltyan@mail.tsinghua.edu.cn}
\affiliation{Key Laboratory of Advanced Materials (MOE), Department of Chemical Engineering, Tsinghua University, Beijing 100084, P. R. China}

\date{\today}

\begin{abstract}
By developing a molecular dynamics model of bacterial chemotaxis, we present the first investigation of tracer statistics in bacterial suspensions where chemotactic effects are considered. We demonstrate that the non-Gaussian statistics of full-coated tracer arises from the athermal bacterial noise. Moreover, Janus (half-coated) tracer performs a composite random walk combining power-law-tail distributed L\'{e}vy flights with Brownian jiggling at low coating concentration, but turns to an enhanced directional transport (EDT) when coating concentration is high. Unlike conventional self-propelled particles, upon increasing coating concentration, the direction of EDT counterintuitively reverses from along to against the tracer orientation. Both these transitions are identified to be second-order, with the phase boundaries meeting at a triple point. A theoretical modeling that reveals the origin of such anomalous transport behaviors is proposed. Our findings reveal the fundamental nonequilibrium physics of active matter under external stimuli, and underscore the crucial role of asymmetrical environment in regulating the transport processes in biological systems.
\end{abstract}

\maketitle

Active responses to external stimuli in living systems, which help organisms to realize adaptive behaviors in changing environments, ubiquitously affect the inner working and collective motion of complex biological fluids. Such responses of active matter can reconstruct the intrinsic irreversible nature of corresponding off-equilibrium dynamics, thereby leading to the emergency of intriguing and complex spatiotemporal behaviors. Like the magnetoreception in bird migration for acclimatization to climate \cite{Hiscock}, and L\'{e}vy walks in animal movements for optimal foraging \cite{Edwards}, the transport kinetics of such active fluids with stimulus responses allows a pivotal approach to reveal the novel statistical physics of nonequilibrium systems.

On the other hand, as a typical active system with net incoming flux of energy, the suspensions of motile bacteria demonstrate a rich variety of intriguing nonequilibrium phenomena such as targeted delivery \cite{Koumakis}, ratchet motors \cite{Angelani,Leonardo} and violation of fluctuation-dissipation theorem \cite{Chen}. Thus, exploring the exotic features and potential applications of such active media is of great physical and biological importance. One reliable and universal approach to probe their inherent properties is analyzing the diffusion dynamics of a passive tracer immersed in bacterial bath \cite{Chen,Wu,Leptos,Wilson,Maggi,Peng}. Rather than in conventional equilibrium media, anomalous behaviors of tracer such as enhanced transport \cite{Wu} and non-Gaussian statistics \cite{Leptos} are widely observed, thereby probing the intrinsic peculiarities of bacterial suspensions. However, chemotaxis \cite{Berg,Endres,Kalinin}, the biased movement of bacteria in response to chemical stimuli, has not been addressed in those previous works. As chemotaxis is a basic and universal phenomenon among motile bacteria, the nonequilibrium physics of such active matter remains unrevealed,
thereby making it an urgent and striking issue to explore the tracer statistics in bacterial suspension with chemotactic effects.

In this Letter, we report the first investigation of tracer statistics in bacterial suspensions where chemotactic effects are considered. A model of bacterial chemotaxis is developed, to study the diffusion dynamics of an isolated chemoattractant-coated spherical tracer in bacterial suspension. As shown in Fig. 1(a), we consider a suspension in a 2D box \cite{Angelani,Maggi,Julian,Redner,Ni,Rein} $L \times L$ ($L=160\mu {\rm m}$) with periodic boundary conditions, consisting of $N$ bacterial cell and a passive spherical tracer with radius $R$. Each cell is represented by a spherocylinder \cite{Angelani}, with a diameter $d=1.5\mu {\rm m}$ \cite{Berg2}, a length $2d$ and an orientation ${\bm e}_i$ denoting the swimming direction (inset). Fig. 1(b) shows the tracers used in our simulations, with the coated portion colored in orange. The motion of bacteria [Fig. 1(c)] is modeled based on the run-and-tumble dynamics of {\it E. coli} \cite{Berg2} and chemotactic responses are realized by Monod-Wyman-Changeux model \cite{Endres,Kalinin}. The details of model are given in Supplemental Material \footnote{ See Supplemental Material for additional methods, movies, and figures.}. Next, to obtain the chemoattractant-concentration field $c$ generated by the coated portion, we assume that chemical spreads in the bath by classical diffusion with diffusion constant $D$, and also decays with a rate $k$ in consequence of the enzymatic activity of bacteria. Therefore, $c$ obeys the reaction diffusion equation \cite{Eshel,Grima,Taktikos}
\begin{equation}\label{rde}
    \partial_t c({\bm r},t) = D\nabla^2c-kc.
\end{equation}
We use $D=500\mu{\rm m^2/s}$ based on the diffusivity of MeAsp \cite{Lazova}, a widely used chemoattractant in experiments \cite{Endres,Kalinin,Lazova}, and $k=10{\rm s}^{-1}$ \cite{Taktikos}. Particularly, the time-dependent solution of Eq. \eqref{rde} has a high convergence rate \footnotemark[\value{footnote}], leading to a rational assumption that $c$ is instantaneously stationary, i.e., $\partial_t c\equiv0$.

\begin{figure}[t]
\includegraphics[width=8.7cm]{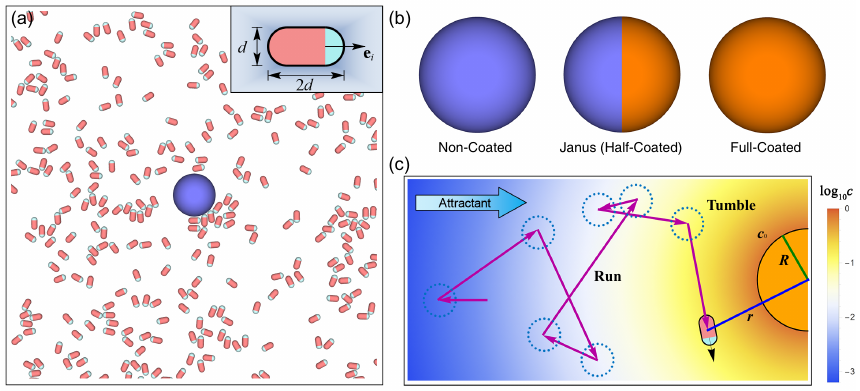}
\caption{\label{fig:epsart}(a) The snapshot of simulation system, where the bacterial cell is represented by spherocylinder as depicted in inset. (b) Three types of tracer used in our simulations. The coated and non-coated portions are colored orange and blue respectively. (c) Schematic diagram of the motion of run-and-tumble bacteria in a chemoattractant-concentration field $c$ generated by the coated portion.}
\end{figure}

Next, to test the validity of this model, we focus on the concentration field for full-coated tracer given by $c(r)=c_0Re^{-\sqrt{k/D}(r-R)}/r$ \footnotemark[\value{footnote}], where $c_0$ is the coating concentration and $r$ is the center-to-center distance between cell and tracer. In particular, the minimal detectable concentration $c^*$ of MeAsp for {\it E. coli} is $c^* = 0.001$mM \cite{Sourjik}, leading to a signal-detectable region (SDR) in which the cell can perform chemotactic behaviors. We specialize to the case of $R = 5.07\mu {\rm m}$ \cite{Wu,Leptos} and $N = 784$ with number density $\phi = N/L^2=0.03 {\rm / \mu m^2}$. As shown in Fig. 2(a) where $c_0 = 1.0$mM, the distribution of bacterial density $\rho$ indeed proves the existence of SDR. Further, the radius of SDR, $r_s$, is measured for different $c_0$ and shown in the inset of Fig. 2(b). The high agreement between $r_s$ and analytic values (red line, determined by $c(r_s)=c^*$) corroborates that chemotactic effects can indeed be reproduced by our model (See also Fig. S3 and Movie I).

Based on this model, the diffusion of full-coated tracer in bacterial suspension is firstly investigated. Average over 20 independent runs are performed for each $c_0$. Fig. 2(a) shows the mean square displacement (MSD) $\langle \Delta x^2(t) \rangle = \langle | {\bm x}(t)-{\bm x}(0) |^2 \rangle$ for a wide range of $c_0$, where ${\bm x}(t)$ is the position vector of tracer and $\langle \cdots \rangle$ denotes the ensemble average. It can be found that tracer motion is short-time superdiffusive and long-time Fickian. Moreover, 2D non-Gaussian parameter $\alpha_2(t)=\langle \Delta x^4(t) \rangle/2\langle \Delta x^2(t) \rangle^2-1$ \cite{Kim} is also calculated to characterize the heterogeneity of tracer statistics [Fig. 2(c)]. The dramatic departure of $\alpha_2(t)$ from 0 at short times shows the emergence of non-Gaussian statistics and a violation of central limit theorem (CLT). One perspective of such violation is that the non-Gaussianity can be attributed to the athermal environmental noise which is strongly correlated \cite{Kanazawa}. To rationalize this perspective in our system, we use the Langevin formalism to describe the motion of tracer:
\begin{equation}
    {\dot x}(t) = \xi_B(t) + \xi_T(t),
\end{equation}
where {\it bacterial noise} $\xi_B$ denotes the perturbations exerted by bacteria and $\xi_T$ is the Gaussian white noise. The autocorrelation functions of $\xi_B$ are calculated, and are found to follow an exponential decay [Fig. 2(d)]. Such correlation is same with the well-known Ornstein-Uhlenbeck (OU) process ${\dot z}(t)=\xi_{OU}(t)$, where noise $\xi_{OU}$ is also exponentially correlated but Gaussian distributed \cite{Romero}. However, the probability distribution $p(z, t)$ for OU process has a Gaussian form \footnotemark[\value{footnote}], quite different from the non-Gaussian tracer statistic (Fig. S4). Therefore, unlike $\xi_{OU}$, $\xi_B$ must be a non-Gaussian (athermal) noise, as predicted by the aforementioned perspective. We further confirm the non-Gaussianity of $\xi_B$ [inset of Fig. 2(e)]. Corresponding non-Gaussian parameter defined as $K_2=\langle \xi_B^4 \rangle/3\langle \xi_B^2 \rangle^2-1$ is also calculated. As shown in Fig. 2(e), $K_2$ and the maximum of $\alpha_2(t)$, $\alpha_{2 \rm max}$, which characterizes the non-Gaussian level of tracer statistics, show same dependence on $c_0$. Such strong correlation between $K_2$ and $\alpha_{2 \rm max}$ corroborates that the non-Gaussian statistics of tracer indeed originates from the athermal bacterial noise.

\begin{figure}[b]
\includegraphics[width=8.7cm]{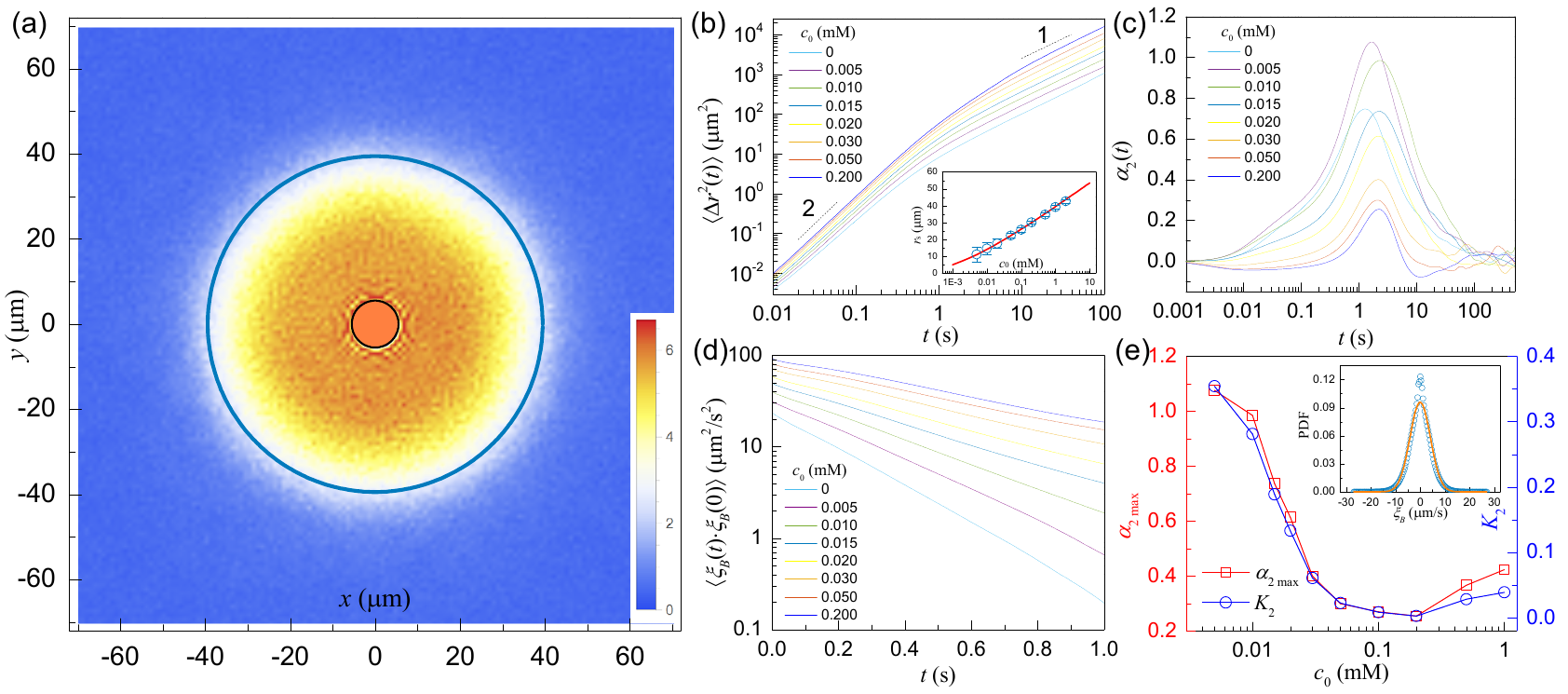}
\caption{\label{fig:epsart}(a) Bacterial density distribution where $c_0=1.0$mM. The blue line denotes the corresponding analytic SDR. (b) Mean square displacements for a wide range of $c_0$. Inset: The radius of SDR $r_s$ versus $c_0$. The red line represents the analytic values. (c) Non-Gaussian parameter $\alpha_2(t)$. (d) Autocorrelation functions of bacterial noise $\xi_B$. (e) The plots of $\alpha_{\rm 2 max}$ and $K_2$ versus $c_0$. Inset: Distribution of $\xi_B$ when $c_0$ = 0.005mM (circle). The solid line is the gaussian distribution with same variance.}
\end{figure}

In striking contrast to full-coated tracer where the concentration field is spherically symmetrical, breaking the symmetry of chemotactic signal can directly lead to a biased bacterial noise ($\langle \xi_B \rangle \neq 0$), thereby complicating the tracer fluctuation. Thus, to quantitatively capture the potential peculiarities of tracer statistics brought by asymmetrical effects, we investigate the diffusion dynamics of a Janus (half-coated) tracer immersed in bacterial bath. By solving Eq. (1), corresponding concentration field can be expressed as \footnotemark[\value{footnote}]
\begin{equation}
    c(r,\theta)=c_0\sum\nolimits_{n = 0}^\infty {a_nk_n(\lambda r)P_n(\cos\theta)},
\end{equation}
where $P_n(t)$ and $k_n(t)$ ($n=0,1,2,3,\ldots$) are the Legendre polynomial and modified spherical Bessel function of the second kind, respectively. The definition of $\theta$ is delineated in the top inset of Fig. 3(a). $a_n$ is determined by boundary condition, given by $a_n=\frac{2n+1}{2k_n(\lambda R)}\int_0^1{P_n(t)dt}$ where $\lambda = \sqrt{k/D}$.

\begin{figure}[b]
\includegraphics[width=8.7cm]{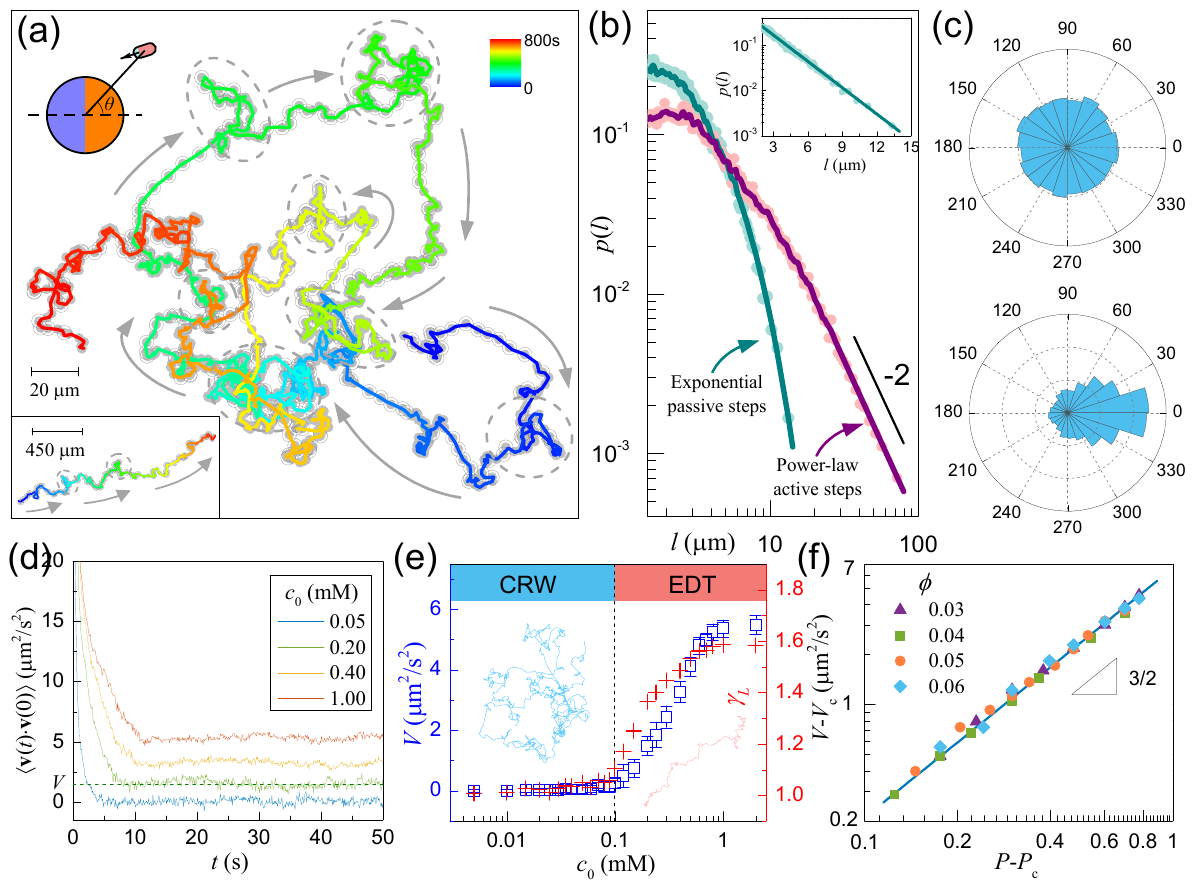}
\caption{\label{fig:epsart}(a) Representative trajectories tracked for 800s when $c_0 = 0.04$ (main view) and 1.0mM (bottom inset) respectively. The top inset denotes the definition of $\theta$. (b) The distribution of step size $l$ plotted in log-log scale. (c) Distributions of turning angles for $c_0 = 0.04$ (top) and $1.0$mM (bottom). (d) Time evolution of VACFs for various $c_0$. (e) The plot of $V$ and $\gamma_L$ versus $c_0$. The dashed line denotes the critical point. (f) The log-log plot of $V-V_c^c$ versus $P-P_c^c$ for various bacterial densities $\phi$.}
\end{figure}

We focus on the system where $\phi = 0.03{\rm / \mu m^2}$ as well. Unlike the motion of full-coated tracer which is always long-time Brownian, Janus tracer demonstrates two distinct transport patterns for different $c_0$. Representative trajectories of such two patterns are shown in Fig. 3(a) for $c_0 = 0.04$ (main view) and 1.0mM (bottom inset) respectively. For low coating concentration (e.g., $c_0=0.04$mM), the tracer alternatively performs persistent walks and local jiggling, which are schematically indicated by arrows and dashed circles respectively. To reveal the inner nature of this unusual randomness, a wavelet-based method \cite{Kejia} is employed to rapidly separate the persistent ``active" runs [highlighted by grey circles in Fig. 3(a)] from random ``passive" jiggling. To uncover the underlying differences between these active and passive steps, we denote the length of active or passive segment as ``step size", $l$, and measure the distributions $p(l)$ based on over 100 independent trajectories. As shown in Fig. 3(b), the size of passive step is exponential distributed (highlighted in inset), indicating that statistics of local jiggling obeys the descriptions of classic Brownian random walks \cite{Codling}. More intriguingly, for active steps, a power-law tail with slope $\mu = -2$ can be identified. Such power-law tail suggests that active steps follow the landscape of L\'{e}vy flights, where the distribution of step size is heavy-tailed with slope $\mu$ satisfying $-3<\mu<-1$ \cite{Plank}. Therefore, the transport of Janus tracer at low $c_0$ is identified to be a composite random walk (CRW) combining power-law-tail distributed L\'{e}vy flights with Brownian jiggling. However, such CRW doesn't hold for high $c_0$ (e.g., $c_0 = 1.0$mM), where the tracer motion maintains directional persistence as delineated in the bottom inset of Fig. 3(a). Such reinforcing directionality can be quantitatively characterized by the distribution of {\it turning angles}, which are determined by an error-radius analysis \cite{Kejia2}. As shown in Fig. 3(c), the uniformity of turning angles for $c_0 = 0.04$mM (top) implies that no direction is preferred for turns in CRW, while the biased distribution for $c_0 = 1.0$mM (bottom) shows that the corresponding transport is statistically directional. Hence, a ``phase" transition for Janus tracer can be identified, from CRW to enhanced directional transport (EDT) when increasing the coating concentration.

Moreover, as demonstrated by the long-time diffusion exponent $\gamma_L$ [Fig. 3(e), red crosses], which is determined by $\langle \Delta r^2(t)\rangle \sim t^{\gamma_L}$ for long time scales, the long-time transport of Janus tracer is found to be Fickian for CRW and superdiffusive for EDT. Corresponding critical point $c_0^c$ can be identified at $c_0 = 0.1$mM. Note that $\langle \Delta r^2(t)\rangle=\int_0^t{ds(t-s)\langle{\bm v}(s)\cdot{\bm v}(0)\rangle}$ where $\langle{\bm v}(t)\cdot{\bm v}(0)\rangle$ is the velocity autocorrelation function (VACF). Hence, as a more precise approach to identify the pattern of such CRW-to-EDT transition (CET), we calculate the VACFs of tracer for various $c_0$ and show them in Fig. 3(d). Over sufficiently long times, VACF decays to zero for $c_0 < c_0^c$, and demonstrates a non-zero steady-state value (denoted as $V$) for $c_0 > c_0^c$. As expected, $V$ and $\gamma_L$ show the same dependence on $c_0$ [Fig. 3(e)], corroborating that such approach can indeed capture the inner change of statistics when undergoing the phase transition. Particularly, the dependence of $V$ upon $\log_{10}c_0$ is continuous-like, indicating such transition may be second-order. A central feature of second-order phase transition is that, physical quantities show power-law dependence with characteristic critical exponents near critical point \cite{Doron}. Hence, to probe this idea in our system, we look for an analogous critical behavior near $c_0^c$, expressed as $V-V_c^c \propto (P-P_c^c)^\beta$,
where $P = \log_{10}c_0$ and $\beta$ is the critical exponent. $V_c^c$ and $P_c^c$ are corresponding values for $c_0^c$. As shown in Fig. 3(f), $V-V_c^c$ versus $P-P_c^c$ demonstrates a power-law relation with critical exponent $\beta = 1.5$. Moreover, since critical exponent is usually system-independent \cite{Yeomans}, we systematically change the bacterial density $\phi$ to examine the universality of such behavior. As expected, power-law relation can still be identified, with $\beta$ remaining 1.5 regardless of $\phi$. Therefore, CET indeed conforms to the scenario of second-order phase transition.

On the other hand, Janus tracer follows the landscape of active self-propelled particle when it undergoes the EDT. For conventional self-motile Janus motors such as catalytic Janus colloids \cite{Palacci} and laser-heated metal-capped particles \cite{Volpe}, the propulsion direction is predetermined by corresponding surface geometry. However, the direction of EDT is not prescribed. Instead, as denoted by the arrows in Figs. 4(a) and (b) where $c_0 = 0.05$ and 1.0mM respectively, upon increasing $c_0$, the propulsion direction undergoes an unexpected transition from along to against the tracer orientation {\bf n} [defined in the inset of Fig. 4(c)]. Such counterintuitive behaviors can be attributed to the asymmetric distribution of bacterial density $\rho$ [Figs. 4(a) and (b)], which directly leads to a bias of bacterial noise. Such bias can be quantitatively described by the non-uniform distribution of included angle $\varphi$ between {\bf n} and instantaneous $\xi_B$. For $c_0 = 0.05$mM, the probability distribution $p(\varphi)$ [Fig. 4(c)] demonstrates that more collisions are exerted on coated side (i.e., $\varphi\in[0,\pi/2]$) due to that chemotactic aggregation only exists near the coated portion [Fig. 4(a)]. Hence, the propulsion is along the direction of {\bf n}. However, for $c_0 = 1.00$mM [Fig. 4(d)], the bias of $\xi_B$ reverses since $\rho$ near non-coated side is higher than the other [Fig. 4(b)], thereby leading to a counterintuitive transport where the direction is against {\bf n}.

To quantify such propulsion direction transition (PDT), we define $v_p=\langle \xi_B\cos\varphi \rangle$ as the propulsion velocity and show the corresponding dependence on $c_0$ in Fig. 4(e) where $R=6.97 \mu{\rm m}$. The sign of $v_p$ denotes the transport direction of tracer. Positive and negative values represent directions which are along and against {\bf n} respectively. Particularly, as shown in the inset of Fig. 4(e), power-law relation can still be identified near the critical point $c_0^p = 0.28$mM, suggesting that such transition is also second-order. Further, to examine the generality of this counterintuitive phenomenon, we walk through the parameter space ($c_0-R$) that is physically realistic and give the phase diagram of $v_p$ in Fig. 4(f). The critical points for CET and PDT are depicted as blue circles and red triangles respectively. It can be found that CET becomes indistinctive when $R$ increases, since more effective collisions can be exerted on tracer with larger size. Moreover, two phase boundaries meet at a triple point near (0.1mM-$5\mu {\rm m}$). That is, PDT only exists for $R > 5\mu {\rm m}$. For tracer with $R < 5\mu {\rm m}$, the transport direction is always counterintuitively against the orientation.

\begin{figure}[t]
\includegraphics[width=8.7cm]{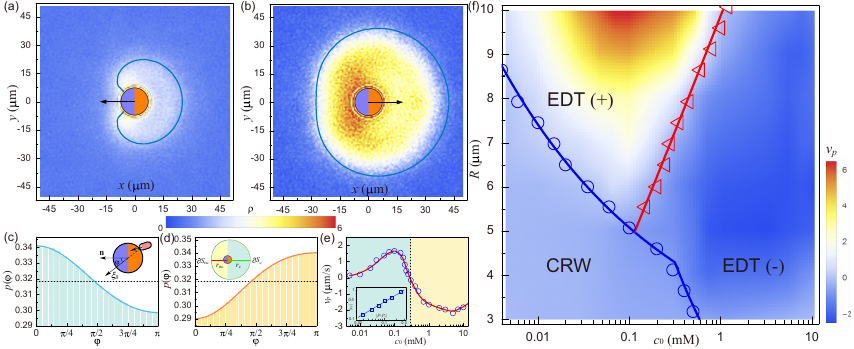}
\caption{\label{fig:epsart}(a)-(b) The distributions of bacterial density $\rho$ for tracer with $R = 6.97\mu m$ where $c_0 = 0.05$mM (a) and 1.0mM (b) respectively. The black arrows denote the corresponding propulsion directions, and the blue lines highlight the corresponding SDRs. (c)-(d) The distributions of $\varphi$ where $c_0 = 0.05$mM (c) and 1.0mM (d) respectively. The dashed lines show the distribution of $\varphi$ without chemotactic effects. The definitions of $\varphi$, $r_{n/nc}$ and $\partial S_{n/nc}$ are shown by the insets. (e) The dependence of $v_p$ on $c_0$ for tracer with $R = 6.97\mu m$. The inset shows the power-law relation between $|v_p|$ and $|P-P_c^p|$, where $P = {\rm log}_{10}c_0$ and $P_c^p = {\rm log}_{10}c_0^p$. (f) The phase diagram of $v_p$ on $c_0-R$ space. The blue circles and red triangles are the critical points for CET and PDT respectively. The corresponding solid lines are analytic phase boundaries determined by Eq. \eqref{pb}.}
\end{figure}

To understand the inner physics of such intriguing phase behaviors, from the perspective of quantifying the biased bacteria noise, we denote $\Phi_c$ and $\Phi_{nc}$ as the net influx of bacteria into the SDR $S$ on the coated and non-coated sides respectively, given by $\Phi_{n/nc}=\int_{\partial {S_{c/nc}}}{j_{c/nc}^{in}\cdot ds}$, where $j_{c/nc}^{in}$ represent the net rates of bacterium gain per unit of region boundary $\partial S_{c/nc}$ [depicted in inset of Fig. 4(d)]. We use the influx difference $\zeta=\Phi_c-\Phi_{nc}$ to characterize the bias level of bacterial noise $\vartheta$. Since CET and PDT are both second-order, $\vartheta$ can be estimated by $\zeta$ using a first-order Taylor series expansion, given by
\begin{equation}
    \vartheta = \kappa_0 + \kappa_1\zeta + \mathcal{O}(\zeta^2),
\end{equation}
where $\kappa_0$ is the intrinsic bias induced by the tracer geometry and $\kappa_1$ is the factor measuring the effective influx which collides with the tracer. For spherical tracer used in our systems, $\kappa_0=0$ and $\kappa_1\propto R$.

In such a way, CET occurs when $\vartheta$ exceeds the maximum bias that can finally be dissipated. Considering that $c_0^c$ for different $R$ are all not high enough, we obtain $\Phi_{nc}\simeq 0$ due to that no chemotactic aggregation is near the non-coated portion. Moreover, in consequence of the small concentration gradient, $j_c^{in}$ can be approximated as constant, thereby making $\Phi_c\propto\int_{\partial {S_c}}{ds}=l_s$, where $l_s$ is the length of $\partial S_c$. Owing to the semicircle geometry of $\partial S_c$, we denote $r_c$ as the distance satisfying $c(r_c,1) = c^*$, then we obtain $l_s \simeq \pi r_c$. Hence, $\vartheta=\kappa_1\zeta\propto Rr_c$, indicating that $c_0^c$ for different $R$ locate at the contour line of $Rr_c$. However, the effects of $\Phi_{nc}$ is nonnegligible for PDT where the boundary is given by $\vartheta<0$, i.e., $\Phi_{nc}>\Phi_c$. To quantify such effects, we denote $r_{nc}$ as the distance satisfying $c(r_{nc},-1)=c^*$. Since $r_{nc} \ll r_c$ at $c_0^p$, the bacterial cells near $\partial S_{nc}$ can reach the tracer faster than near $\partial S_{c}$, thereby making $\Phi_{nc}>\Phi_c$ when $S_{nc}$ is large enough to allow chemotactic aggregation. As the effective size of $S_{nc}$ can be characterized by $r_{nc}/R$, PDT will occur once $r_{nc}/R$ reaches the threshold value. Hence, the analytic phase boundaries for two transitions read
\begin{equation}\label{pb}
    Rr_c = \chi_c,~r_{nc}/R = \chi_{nc},
\end{equation}
where the values $\chi_c$ and $\chi_{nc}$ are obtained using critical points (0.10mM-5.07$\mu {\rm m}$) and (0.28mM-6.97$\mu {\rm m}$) respectively. As shown in Fig. 4(f), both phase boundaries determined by Eq. \eqref{pb} fit the simulation results well, corroborating that the origin of such counterintuitive transport behaviors can be theoretically revealed from capturing the symmetry breaking of bacterial noise.

In summary, by developing a model of bacterial chemotaxis, we have presented what is to our knowledge the first investigation of tracer statistics in bacterial suspensions with chemotactic effects. We demonstrate that the non-Gaussian statistics of full-coated tracer originates from athermal bacterial noise. Moreover, Janus tracer performs CRW and EDT for low and high coating concentrations respectively. Unlike conventional self-propelled particles, upon increasing the coating concentration, the direction of EDT undergoes a counterintuitive transition from along to against the tracer orientation. Both these transitions are identified to be second-order, with the phase boundaries meeting at a triple point. A theoretical modeling is proposed from the perspective of quantifying the symmetry breaking of bacterial noise, which reveals the origin of such anomalous transport behaviors. Our findings provide a significant advance in revealing the fundamental nonequilibrium physics of active matter under external stimuli, and suggest a novel approach for efficient cargo delivery utilizing stimulus-response technique and asymmetrical design.

\begin{acknowledgments}
We are thankful for helpful disscusions with Ye Yang and Tianqi Cui. We acknowledge financial support from National Natural Science Foundation of China (Grant Nos. 21422403, 51273105, 51633003, 21174080). L.-T. Y. acknowledges financial support from Ministry of Science and Technology of China (Grant No. 2016YFA0202500).
\end{acknowledgments}

\bibliography{Bibliography}

\end{document}